\documentclass{tMPH2e}
\usepackage{graphicx}

\title{Simulation-based equation of state of the hard disk fluid\\ and prediction of higher-order virial coefficients}

\author{J. KOLAFA$^*$\thanks{$^*$Corresponding author. Email:
jiri.kolafa@vscht.cz} and M. ROTTNER\\Department of Physical
Chemistry, Prague Institute of Chemical Technology, 166 28 Praha 6,
Czech Republic}

\begin{document}
\markboth{J. Kolafa and M. Rottner}{Hard disk fluid}

\maketitle

\begin{abstract}
We present new molecular dynamics results for the pressure of the pure
hard disk fluid up to the hexatic transition (about reduced density
0.9).  The data combined with the known virial coefficients (up to
$B_{10}$) are used to build an equation of state, to estimate
higher-order virial coefficients, and also to obtain a better value of
$B_{10}$.  Finite size effects are discussed in detail.  The ``van der
Waals-like'' loop reported in literature in the vicinity of the
fluid/hexatic transition is explained by suppressed density
fluctuations in the canonical ensemble.  The inflection point on the
pressure-density dependence is predicted by the equation of state even
if the hexatic phase simulation data are not considered.
\end{abstract}


\section{Introduction}

The hard disk fluid is the simplest model of various surface systems
(surfactants, adsorption of molecules on smooth surfaces).  It has
been intensively studied by a number of methods, of which we are
interested in Monte Carlo and molecular dynamics simulations
\cite{EW1,EW2,Mak,Jaster}, virial coefficients
\cite{virials,virialsLKM}, and equations of state (EOS)
\cite{DP19,DP20,DP21,DP22,DP23,DP24,DP25,DP26,DP27,DP28,DP29,Eisenberg}.
Recently, peculiarities of phase transitions from fluid to hexatic to
solid are frequently studied \cite{Eisenberg,Jaster}.

In this work we focus mainly on the fluid region.  The fluid phase
changes at density of about $\rho_{\rm c}=0.8995(23)$ ($\rho$ denotes
the reduced number density, $\rho=N\sigma^2/A$, where $N$ is the
number of disks, $\sigma$ their diameter, and $A$ the system area)
to the hexatic phase \cite{Binder,JasterOO,Eisenberg}.

\section{Methods}

\subsection{Simulation methodology}

\subsubsection{Molecular dynamics code}

To obtain accurate data on the EOS, we use standard
molecular dynamics (MD) simulations in the microcanonical ensemble and
tetragonal (square) periodic boundary conditions with zero total
momentum (MD-NVE; we keep the usual abbreviation NVE for constant
Number of particles--Volume--Energy even if the volume is here
replaced by the area).
Our MD program combines the ideas of the linked-cell list method both
in space and time \cite{EW1,EW2} and is highly optimized.  Details on
the code are given in our previous paper \cite{hsmd}.  In the present
study, we use $N=4000$ and 9000, and at higher densities 16\,000, 25\,000
and 50\,000 disks of unit diameter in a square periodic box.

\subsubsection{Compressibility factor}

The compressibility factor $Z=pA/(NkT)=p/(\rho kT)$ ($p$ denotes
pressure,  $k$ the Boltzmann constant, and $T$
absolute temperature) can be calculated from MD  simulations in two
ways \cite{EW1,EW2}.  One formula uses the virial of force 
\begin{equation}
Z_{\rm vir}(t_1,t_2) = 1 - {N-1\over N}\,
 {1\over  2 E_{\rm kin} (t_2-t_1)}
 \sum_{t\in(t_1,t_2)} \Delta {\bf v}_{ij} \cdot {\bf r}_{ij} 
,
\label{Zvir}
\end{equation}
where the sum is over all collisions occurring during time interval
$(t_1,t_2)$, $\Delta {\bf v}_{ij}$ is the change of velocity of both
colliding disks of mutual position ${\bf r}_{ij}$, and $E_{\rm kin}$
is the kinetic energy (constant in the MD-NVE simulation).  The
alternative formula uses the collision rate
\begin{equation}
  Z_{\rm rate}(t_1,t_2) 
 = 1 + \gamma(N) \sqrt{\pi\over 2D N E_{\rm kin}}\;
 {1\over t_2-t_1}  \sum_{t\in(t_1,t_2)} 1
,
\label{Zrate}
\end{equation}
where $\sum_{t\in(t_1,t_2)} 1$ is the number of collisions in
time interval $(t_1,t_2)$ and \cite{EW2}
\begin{equation}
\gamma(N) = {\Gamma[(D(N-1)+1)/2] \over \Gamma[D(N-1)/2] (DN/2)^{1/2}},
\end{equation}
where $D=2$ is the dimensionality.

\begin{figure}
\begin{center}\includegraphics[scale=0.75]{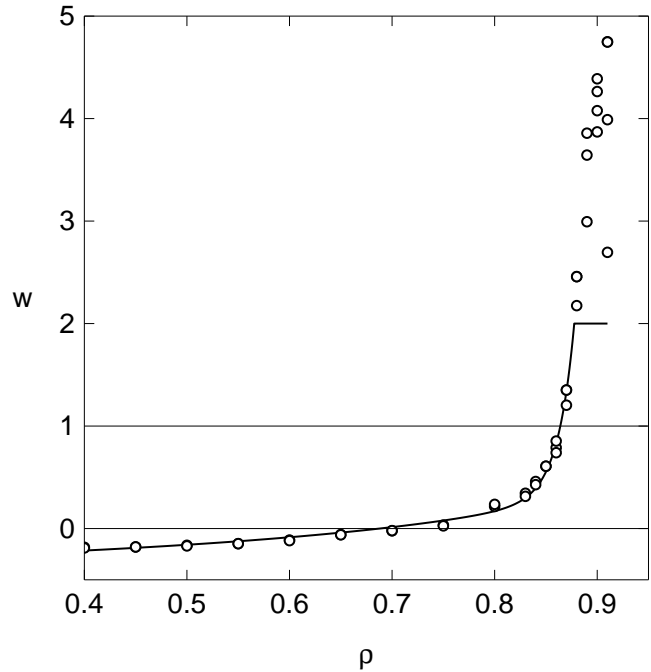}\end{center}
\caption{Weights $w$ for mixing both formulas for $Z$ in equation
(\ref{ZMD}) and the fit bounded by $w\le 2$.}
\label{fig1}
\end{figure}

Both formulas give within statistical inaccuracies the same values,
but their statistical errors differ.  Our final result is therefore a
weighed average of both formulas,
\begin{equation}
  Z = w Z_{\rm vir} + (1-w) Z_{\rm rate}  ,
\label{ZMD}
\end{equation}
where the mixing weight $w=w(\rho)$ is a function of density.
Methodology for determining the optimum function $w$ has been
explained in detail in \cite{hsmd}. The simulation results on
$w(\rho)$ are shown in figure~\ref{fig1} along with a function fitted
to the data and bounded by $w\le 2$ to avoid too
extrapolated values.  This function was used in final analysis of MD
data to avoid bias.

\subsubsection{Start and equilibration}

The initial configuration was a tetragonal crystal with random
velocities assigned to particles.  This setup creates a disorder
within a few collisions.  At higher densities, a locally hexagonal
arrangement gradually develops.  Therefore a period of equilibration
follows until the compressibility factor reaches a constant value with
fluctuations.  At higher densities and for a large system this takes a
long time.

\subsection{Finite size effects}

\subsubsection{Fluid region}

\begin{table}
\tbl{Molecular dynamics results on the compressibility factor,
  corrected for the finite size effects; for $\rho<0.88$ by equation (\ref{Zenscorr}), for $\rho>0.88$ by $Z(1/N)$ linear extrapolation.}
{\begin{tabular}{lll}\toprule
$~\rho$ & $~~Z$ & $~~\sigma(Z)$ \\\colrule
    0.40 &  2.1514393 & 0.0000052 \\
    0.45 &  2.4276680 & 0.0000065 \\
    0.50 &  2.7601235 & 0.0000079 \\
    0.55 &  3.1647878 & 0.0000098 \\
    0.60 &  3.663691  & 0.000012 \\
    0.65 &  4.287926  & 0.000015 \\
    0.70 &  5.082362  & 0.000019 \\
    0.75 &  6.113391  & 0.000026 \\
    0.80 &  7.476491  & 0.000036 \\
    0.83 &  8.494891  & 0.000050 \\
    0.84 &  8.866011  & 0.000059 \\
    0.85 &  9.245785  & 0.000072 \\
    0.86 &  9.621609  & 0.000082 \\
    0.87 &  9.96782   & 0.00013  \\
    0.88$^{\rm a}$&10.2309  & 0.0003 \\  
    0.89 & 10.3176    & 0.0011 \\ 
    0.90$^{\rm b}$ & 10.2059    & 0.0011 \\ 
\botrule
\end{tabular}}\label{tab1}
\tabnote{$^{\rm a}$ A compromise between  10.23095(26) by (\ref{Zenscorr}) and
10.23055(54) by $Z(1/N)$ extrapolation.}
\tabnote{$^{\rm b}$ This value may be affected by finite size effects.}
\end{table}

The finite-size errors are in our pseudoexperimental setup at low
densities several times larger than the statistical errors.
E.g., at the lowest density simulated, $\rho=0.4$, one would need as
many as $N=160\,000$ particles to guarantee the same systematic finite-size
error as the statistical error of table~\ref{tab1}.
It is more efficient to simulate a smaller system and to correct for
the finite-size error to obtain the thermodynamic limit
($N\rightarrow\infty$) because correlation times and therefore
convergence in large two-dimensional systems is slow.  At about
$\rho=0.85$ both errors become comparable for $N=20\,000$.
In a vicinity of the hexatic transition the finite-size errors become
large again and difficult to determine.

The correction procedure can be divided into three steps
\cite{hsmd,bridge}:
$$
\mbox{MD-NVE} \longrightarrow \mbox{NVT} \longrightarrow \mu\mbox{VT} \longrightarrow
\mbox{thermodynamic limit},
$$
where the first step is already included in formulas (\ref{Zvir}) and
(\ref{Zrate}), and $\mu$VT stands for the grand-canonical ensemble.

The largest source of inaccuracy for hard-body systems is the second
step, NVT $\rightarrow \mu$VT \cite{yperiodic,korekce}.  This
``ensemble correction'' is caused by suppressed fluctuations of
density (number of particles) in the canonical ensemble and is given
by \cite{hsmd,finitesize}
\begin{equation}
 Z_{\rm \mu VT}-Z_{\rm NVT} =
 {\rho\over2N}\,{\partial^2 p\over\partial\rho^2} \left[\partial p\over
\partial\rho\right]^{-1} + {\cal O}(N^{-2})
,
\label{Zenscorr}
\end{equation}
where indices $\mu$VT and NVT refer to the respective expectation
values in the grand-canonical and canonical ensembles at the same
number density (i.e., $\langle N \rangle_{\mu\rm VT} = N_{\rm NVT}$).  
This term requires knowledge of an EOS which is not
known in advance.  The correction procedure
is therefore self-consistent (we start with no correction, fit the
EOS, calculate the correction and better data, etc.),
albeit rapidly converging.  For $\rho>0.88$, the accuracy of the
EOS is not sufficient to calculate the second derivative needed in
(\ref{Zenscorr}) and we use linear extrapolation of $Z$ in dependence
on $1/N$ to zero.  In a close vicinity of $\rho_{\rm c}$ even this
approach fails.

\subsubsection{Near the critical point}

It was investigated in detail in \cite{Eisenberg} that the $p(\rho)$
dependence has an inflection point at the fluid/hexatic transition,
$p=ZkT\rho=p_0-{\rm const}\times(\rho_{\rm c}-\rho)^{\alpha'}$ as
$\rho\to\rho_{\rm c}$, where $\alpha'=4.55$.
Then the ``ensemble error'' (of opposite sign than the ensemble
correction) becomes
\begin{equation}
 Z_{\rm NVT}-Z_{\rm \mu VT} = {\alpha'-1\over\rho_{\rm c}-\rho}\,{\rho\over 2N}
\label{criterr}
\end{equation}
and similarly for $\rho>\rho_{\rm c}$ (likely with a different
critical exponent $\alpha'$).  This term is positive for
$\rho<\rho_{\rm c}$, negative for $\rho>\rho_{\rm c}$, and it diverges
at $\rho_{\rm c}$.  For $\rho=0.895$ and $N=256^2$ we obtain from
(\ref{criterr}) the NVT pressure by 0.005 higher than the
grand-canonical one which is in qualitative agreement with the value
of 0.01 by \cite{Jaster} (the statistical error is not provided, but
it is likely less than 0.01).  The correct value of the correction may
be affected by higher order terms; apparently this must be true in
a close vicinity of $\rho_{\rm c}$ because the ensemble correction
cannot diverge at finite $N$.  Therefore the ``van der Waals-like''
loop \cite{Mak} is an artifact caused by suppressed density
fluctuations in the constant-$N$ ensemble and it disappears if data
are free of finite-size effects.

Near the first-order phase transition (occuring e.g.\ for hard
spheres) one observes a hysteresis caused by long-lived metastable
states rather than a ``loop''; a loop can arise at very small systems
and for long runs so that an equilibrium is maintained.  There is no
true hysteresis near the continuous transition point, although of
course the dynamics (and convergence) slows down.

\subsection{Equation of state}

\subsubsection{Correlation of the data}

The pseudoexperimental data were fitted to a polynomial in
$x=y/(1-y)$, where $y = A_{\rm HD}\rho$ is the packing fraction and
$A_{\rm HD}$ is the disk area.  The compressibility factor reads as
\begin{equation}
  Z(y) = \sum_{i=0}^k A_i \left(y\over 1-y\right)^i 
 ,
\label{EOS}
\end{equation}
where $A_0$--$A_4$ are determined so that virial coefficients
$B_2$--$B_5$, known either analytically \cite{B2-4,B2-4x} or ($B_5$
\cite{B5}) with high precision, are exactly reproduced.
Parameters $A_i$, $4<i\le k$, are adjustable, but some of them may be
zero.  The number of degrees of freedom (number of all data minus
number of adjustable parameters) is denoted $n_{\rm free}$.

This choice needs some explanation.  Terms $y/(1-y)$ appear in several
theories for both hard spheres and hard disks: the scaled particle
theory \cite{SPT}, Percus-Yevick and hypernetted-chain integral
equations \cite{PY}, and resummation of approximated virial
coefficients (Carnahan-Starling equation) \cite{CS}.  The main reason
for this form is that it moderates the sharp increase of $Z$ with
increasing $y$ as suggested in \cite{Gelbart} (equation (13)) under
name $y$-expansion (function $y$ is here called $x$ and should not
be confused with packing fraction~$y$).

Another popular choice is a rational function which is closely related
to the Pad\'e approximant \cite{Pade} derived from a formal virial
power series; the coefficients are determined by the least-square
method.  There is some experience with this method in our laboratory
\cite{MV}, but detailed investigation including high densities 
found this method numerically unstable and less flexible; we have not
checked this approach for hard disks, though.  In addition, we have
theoretical objections against interpretation of the pole (zero point
of the denominator) of such a rational function as the random close
packing of the hard sphere fluid (frozen glass).  This interpretation
assumes that the EOS can be analytically continued beyond the freezing
point which is not true \cite{nonanal,BinderNA,Ising}; it can be
continued with a limited precision only---the random close packing is
an inaccurate concept in principle.  With increasing precision of
input data this inaccuracy may cause problems in fitting.  In the hard
disk system it is impossible to extrapolate beyond the continuous
transitions (with inflection points on the pressure--density
dependence) to get any random close packing and even this inaccurate
reason for using rational functions becomes invalid.

Another consequence of the nonanalyticity is that the radius of
convergence of the virial series is less than or equal to the phase
transition density (first-order freezing of the hard sphere fluid or
continuous transition of the hard disk fluid).  Consequently the
approximately quadratic $B^*_n$ dependence (for hard spheres) and
linear dependence (for hard disks), related to expanded $y/(1-y)$
terms, cannot extent to infinite $n$.  It is in principle possible to
determine the radius of convergence from the sequence of $B_n$, but
the available precision and the number of terms are not sufficient.

With functional form (\ref{EOS}), the standard objective function
\begin{equation}
s^2 =
{1\over n_{\rm free}}\left\{
 \sum_{j=1}^{n_{\rm data}}
\left[Z(y_j)-Z_j\over\sigma(Z_j)\right]^2
+\sum_{j=6}^{10}
\left[B_j^{\rm EOS}-B_j^{\rm MC}\over\sigma(B_j^{\rm MC})\right]^2\right\}
\label{sigma}
\end{equation}
was minimized, where $\sigma$ stands for the standard error, $B_j^{\rm
EOS}$ is the virial coefficient calculated from the EOS and $B_j^{\rm
MC}\pm \sigma(B_j^{\rm MC})$ are virial MC data with standard errors
\cite{virials,virialsLKM}.  
In other words, both the MD data on the compressibility factor and the
MC data on the virial coefficient are correlated simultaneously.

The value of $s$ for an optimum fit is around unity provided that the
input standard errors $\sigma$ are reliable, which is the case of our
simulations where $\sigma$ is determined with accuracy (error of
error) of a few per cent \cite{hsmd}.  If $s\gg1$ then the number of
adjustable parameters is not sufficient do describe the data.  On the
other hand, one should not use more parameters than necessary because
just noise would be fitted; the best test is to remove one parameter
and to observe whether $s$ significantly increases.

\subsubsection{Higher-order virial coefficients}

Expanding equation (\ref{EOS}) in powers of density gives virial
coefficients.  Virial coefficients $B_i$ for $i\le 5$ are exactly
reproduced, for $6\le i\le10$ they are modified because their change is
allowed by a simultaneous fit (within statistical errors), and for
$i>10$ they are predicted.

\section{Results}

\subsection{Molecular dynamics}

First, we have checked that both the virial (\ref{Zvir}) and rate
(\ref{Zrate}) routes to the compressibility factor are equivalent.
The differences are within combined statistical errors the same and
only for two systems with $N=9000$ and $\rho=0.7$ and
$\rho=0.75$ slightly exceed one standard deviation. 
In fact, because of correlations in the data, the difference is at
low and large densities much less than the combined standard error.

In order to assess the role of the periodic errors, we fitted the total
correlation function $h(r)=g(r)-1$ at large separations to attenuated
oscillations, $h(r)={\rm Re}[A\exp(-Br)/r]$, where $A$ and $B$ are
complex constants and Re denotes the real part.  The value of ${\rm
Re}(B)$ describes the decay of correlations and its typical value at
simulation square size $L=A^{1/2}$, ${\rm Re}(A)\exp[-{\rm
Re}(B)L]/L$, is the estimate of the periodic error.  We found
that ${\rm Re}(B)=0.22$ for $\rho=0.88$.  Consequently the periodic
error is negligible for all used $N$ and $\rho\le0.88$, and therefore
formula (\ref{Zenscorr}) or linear $Z(1/N)$ dependence is sufficient to
account for finite size effects.  For $\rho=0.89$ we found ${\rm
Re}(B)=0.11$ and $N=4000$ may have a small periodic error about
$2\cdot10^{-5}$; the datum was nevertheless discarded.  For
$\rho=0.90 \approx \rho_{\rm c}$ it holds ${\rm Re}(B)=0.044$ and even
$N=16\,000$ is barely sufficient.

For $\rho<0.88$, the data for different $N$ were corrected by
(\ref{Zenscorr}) and a weighed average was taken; a check was made
that the corrected data do match within statistical errors.
Correction term (\ref{Zenscorr}) is not applicable for $\rho\ge0.89$
because it contains the second derivative of the EOS and therefore
the correction term is large and not available with sufficient
precision.  Linear extrapolation of $Z(1/N)$ was used instead; the
final results thus cease precision.  Point $\rho=0.88$ is a
borderline between applicability of both approaches: The ensemble
correction is large and its accuracy may affect the results, but the
extrapolation gives less accurate data.

The final corrected MD data are collected in table~\ref{tab1}.  The
data in the ``difficult'' region close to the phase transition agree
well with recent extensive Monte Carlo data \cite{Jaster,Mak} with the
exception of density $\rho=0.9$ closest to $\rho_{\rm c}$ where
the $N=1024^2$ result $Z=10.212$ \cite{Mak} is significantly larger
than our $Z=10.206$; in this case one cannot assume linearity of the
$Z(1/N)$ dependence and our value is probably affected by higher-order
finite size errors.

\subsection{Equations of state}\label{s:eos}

\begin{figure}
\begin{center}\includegraphics[scale=0.75]{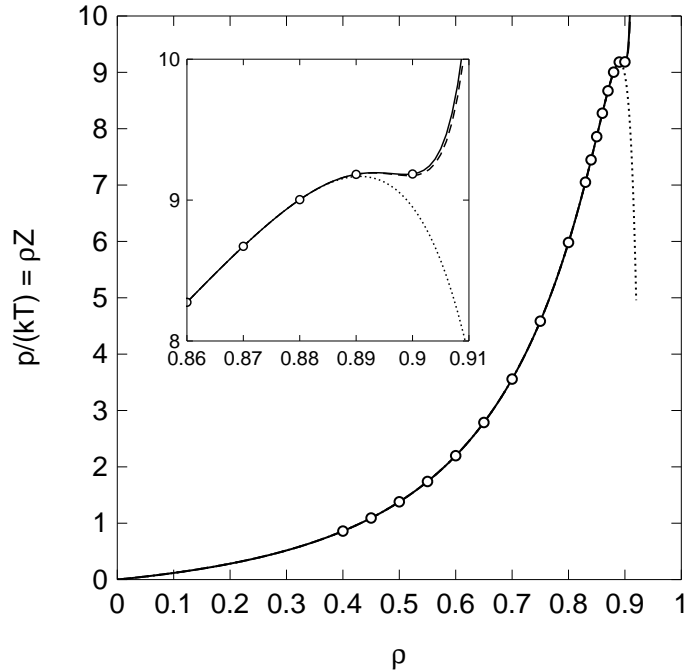}\end{center}
\caption{The data and proposed EOS.  
Dotted line: $\rho_{\rm max}=0.88$, 
dashed line: $\rho_{\rm max}=0.89$,
solid line: $\rho_{\rm max}=0.90$.}
\label{fig2}
\end{figure}

We present three best versions differing by the maximum density
$\rho_{\rm max}$, number of fitted parameters and the value of the objective
function $s$ (\ref{sigma}), see figure~\ref{fig2}.  Note that
$x=y/(1-y)$, where $y$ is the packing fraction.

\bigskip \noindent 
$\rho_{\rm max}=0.88$, $s=0.724$:
\begin{center}
$Z~=~1+2\,x+1.12801775\,x^{2}+0.00181895291\,x^{3}-0.0526134737\,x^{4}+0.0504951668\,x^{5}-0.0325433846\,x^{6}+0.0133946531\,x^{7}+0.00174265604\,x^{8}-0.00944632202\,x^{9}+0.00851111768\,x^{10}-0.0035963525\,x^{11}+0.000577345106\,x^{12}-1.06399127\cdot10^{-7}\,x^{19}$
\end{center}
$\rho_{\rm max}=0.89$, $s=0.966$:
\begin{center}
$Z~=~1+2\,x+1.12801775\,x^{2}+0.00181895291\,x^{3}-0.0526134737\,x^{4}+0.0504963915\,x^{5}-0.0325578581\,x^{6}+0.0134816028\,x^{7}+0.00129187484\,x^{8}-0.00808881628\,x^{9}+0.00669011963\,x^{10}-0.00250795961\,x^{11}+0.000336036442\,x^{12}-5.15282664\cdot10^{-9}\,x^{22}+5.57730095\cdot10^{-23}\,x^{57}$
\end{center}
$\rho_{\rm max}=0.90$, $\sigma=0.927$; region $\rho\in[0.89,0.90]$ of this equation may be affected by finite size effects:
\begin{center}
$Z~=~1+2\,x+1.12801775\,x^{2}+0.00181895291\,x^{3}-0.0526134737\,x^{4}+0.0504960168\,x^{5}-0.0325537792\,x^{6}+0.0134578632\,x^{7}+0.00140888182\,x^{8}-0.00834273601\,x^{9}+0.00694127367\,x^{10}-0.00262254723\,x^{11}+0.000355746352\,x^{12}-5.24672938\cdot10^{-9}\,x^{22}+5.88054639\cdot10^{-23}\,x^{57}$
\end{center}

It is interesting that both equations with $\rho_{\rm max}\ge0.89$
predict to some extent the loop (with the ``classical'' critical
exponent $\alpha'=3$) at the critical (fluid/hexatic) point, even if
this is not the aim of the present work which focuses rather on the
low-density region.  Any extrapolation to $\rho>\rho_{\rm c}$ should
be done with caution because function $p(\rho)$ is likely
nonanalytical at $\rho_{\rm c}$.

\subsection{Virial coefficients}

\begin{figure}
\begin{center}\includegraphics[scale=0.75]{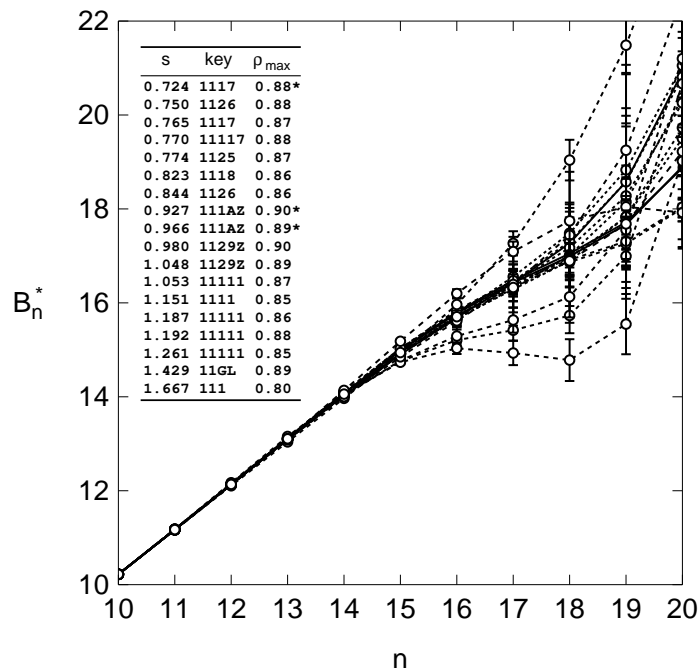}\end{center}
\caption{Higher-order virial coefficients as predicted by different
EOSs.  The best equations from Section \ref{s:eos} are marked by
$\star$ in the table and by solid lines, otherwide the dash size is
proportional to the value of $s$ (the longer dash, the worse fit).
The key determines the equation: a digit (A=10, etc.) stands for one
adjustable parameter (in addition to five parameters corresponding to
$B_6$ to $B_{10}$), its value is the power of $y$ with respect to the
previous term. Error bars are standard errors propagated from standard
errors of the MD data and known virial coefficients. }
\label{fig3}
\end{figure}

The virial coefficients were determined from a number of EOSs, see
figure~\ref{fig3}.  The error of a predicted virial coefficient
consists of a statistical error which is calculated from standard
errors of input data and from a systematic (method) error.  The former
error ranges from typically 0.0014 (max. 0.002) for $B^*_{10}$ to
typically 0.02--0.05 (max. 0.07) for $B^*_{14}$ to typically 0.04--0.2
(max. 0.4) for $B^*_{16}$.  The latter error is apparently difficult
to determine.  We believe that the range of different predictions,
also those with $s>1$, gives a certain measure the systematic error;
it also approximately matches the maximum statistical errors found in the
used set od EOSs.  We are rather pessimistic in determining this
error; most EOSs with $s<1$ (thin lines in figure~\ref{fig3}) give
several times less scattering in the virials predicted and only
moderately increased statistical errors.  All virial coefficients are
collected in table~\ref{tab2}.

\begin{table}
\tbl{Virial coefficients reduced by the disk area (packing fraction
  expansion), $B^*_i=B_i A_{\rm HD}^{1-i}$.  The $B_6$ to $B_9$ values
  are combined data of \cite{virials,virialsLKM}, for $B_{10}$ see
  ref.~\cite{virialsLKM}, $B_{11}$ to $B_{16}$ are our EOS-based 
  predictions with conservatively estimated error bounds (not standard
  errors).}
{\begin{tabular}{lll}\toprule
 $n$ & $B^*_n$ & $\sigma(B^*_n)$ \\\colrule
 2&  2& 0\\
 3&  3.1280177516& 0 (rounded)\\
 4&  4.2578544562& 0 (rounded)\\
 5&  5.33689664& 0.00000064\\
 6&  6.3630259 & 0.0000109 \\
 7&  7.352077  & 0.000028  \\
 8&  8.318677  & 0.000061  \\
 9&  9.27234   & 0.00027   \\
10& 10.2161    & 0.0041    \\\colrule 
10$^{\rm a}$ &10.2203& 0.002     \\\colrule 
10$^{\rm b}$ &10.2210& 0.002     \\ 
11 &11.172     & 0.010     \\ 
12 &12.132     & 0.03      \\ 
13 &13.097     & 0.06      \\ 
14 &14.053     & 0.08      \\ 
15 &14.94      & 0.21      \\ 
16 &15.7       & 0.4       \\ 
\botrule
\end{tabular}}
\tabnote{$^{\rm a}$ EOS-based prediction incl.\ $B_{10}$ of \cite{virials}; the recommended value.}
\tabnote{$^{\rm b}$ EOS-based prediction ($B_{10}$ of \cite{virials} not used).}
\label{tab2}
\end{table}

To verify the procedure, we repeated the calculations with $B_{10}$
removed from the second sum of (\ref{sigma}).  The predicted value was
$B_{10}=10.2210\pm0.002$, which is in agreement with the direct MC
datum \cite{virials} (and in fact more accurate).  The ``best'' value
based on all available data (incl.\ $B_{10}$ of \cite{virials}) is
only slightly smaller, $B_{10}=10.2203\pm0.002$.  In contrast,
lower-order virials are accurate enough and including the MD data in
the fit does not improve precision.

\section{Concluding remarks}

The proposed equations of state in the fluid region combine all
available information---virial coefficients and simulation
compressibility data.  The equations may serve in perturbation theories.
They are not meant as a replacement of physically-based (but less
accurate) equations which are able to described more phases.

The equations enable prediction of higher-order virial coefficients
with no additional assumption on their order-dependence.  The results
also witness about the ``law of complexity conservation'': The value
of the tenth virial coefficient can be obtained with comparable
precision both directly by diagrammatic techniques \cite{virials} and
by simulations.

In order to obtain highly accurate MD data, it was necessary to take into
account finite-size effects, which is especially peculiar close
to the critical fluid/hexatic point.  The ``van der Waals-like'' loop
reported by several authors in this region can be semiquantitatively
predicted by the concept of suppressed density fluctuations in the
canonical ensemble.

\subsection*{Acknowledgments}

This work was supported by the The Ministry of Education, Youth
and Sports of the Czech Republic under the project  LC512
(Center for Biomolecules and Complex Molecular Systems).

This paper was presented at The Seventh Liblice Conference on the
Statistical Mechanics of Liquids (Lednice, Czech Republic, June
11--16, 2006).

\end{document}